\documentclass[a4paper]{spie}  %>>> use this instead for A4 paper
\usepackage{amsmath,amsfonts,amssymb}
\usepackage{graphicx}
\usepackage[colorlinks=true, allcolors=blue]{hyperref}
\usepackage{hyperref}
\title{Calibration strategy for the SPICA/SAFARI instrument}

\author[a]{Russell F. Shipman}
\author[b]{Bart Vandenbussche}
\author[a]{Edgar Castillo-Dominguez}
\author[c]{Alvaro Labiano}
\author[a]{Willem Jellema}
\author[a]{Angiola Orlando}
\affil[a]{SRON, Landlevel 12, Groningen, Netherlands}
\affil[b]{KUL, Leuven, Belgium}
\affil[c]{Centro de Astrobiología (CAB, CSIC-INTA), ESAC Campus, E-28692 Villanueva de la Ca\~nada, Madrid, Spain}

\authorinfo{Further author information: (Send correspondence to R.F, Shipman)\\R.F. Shipman.: E-mail: R.F.Shipman@sron.nl , }

\begin{document} 
\maketitle

\begin{abstract}
SPICA is a mid to far infra-red space mission to explore the processes that form galaxies, stars and planets.  SPICA/SAFARI is the far infrared spectrometer that provides near-background limited observations between 34 and 230 micrometers.  The core of SAFARI consists of 4 grating modules, dispersing light onto 5 arrays of TES detectors per module. The grating modules provide low resolution (250) instantaneous spectra over the entire wavelength range.  The high resolution (1500 to 12000) mode is accomplished by placing a Fourier Transform Spectrometer (FTS) in front of the gratings.  Each grating module detector sees an interferogram from which the high resolution spectrum can be constructed.   SAFARI data will be a convolution of complex spectral, temporal and spatial information.  Along with spectral calibration accuracy of $<$ 1 \%,  a relative flux calibration of 1\% and an absolute flux calibration accuracy of 10\% are required. This paper will discuss the calibration strategy and its impact on the instrument design of SAFARI
\end{abstract}

\keywords{Grating spectrometer, Fourier transform spectroscopy, TES bolometers}

\section{INTRODUCTION}
\label{sec:intro}  % \label{} allows reference to this section
SPICA (SPace Infrared telescope for Cosmology and Astrophysics), one of three remaining candidates for the ESA Medium Class M5 mission and due to compete for the M5 slot through ESA’s Mission Selection Review process starting in early 2021, was unexpectedly cancelled by ESA in October 2020 on financial grounds. With an actively cooled 2.5m-diameter telescope, this space observatory would have comprised three instruments (SAFARI – SpicA FAR-infrared Instrument, SMI – SPICA Mid-Infrared Instrument and B-BOP – magnetic field (B) explorer with BOlometric Polarimeter) working in synergy to provide unprecedented spectroscopic and photometric sensitivity in the mid- and far-infrared, transforming infrared astronomy in fields ranging from planet formation to star formation through to studies of the interstellar medium and galaxy evolution.

This paper/poster is one in a series of SPIE contributions from the SPICA Consortium that is aimed at preserving the technological developments and knowledge gained through work undertaken by many scientists over several years, ensuring that the legacy of SPICA is not lost through the untimely cancellation of this mission. Further details on the SPICA mission are discussed in "The joint infrared space observatory SPICA:  Unveiling the obscured universe"\cite{2020..SPICA.R}. 

%The next generation of far infrared instruments will make use of the technological advances in detector technology of the past decade\cite{Lee:98}\cite{2004IEEEP..92.1597Z}.   
%The advances have been significant and have increased sensitivity, linearity, number of elements and dynamic range. 
The SPICA Far Infrared Spectrometer (SAFARI) will improve on the spectral line sensitivity of previous missions (the Herschel and Spitzer space telescopes) by taking advantage of a cold telescope and unprecedentedly sensitive detectors\cite{SAFARI_det}. Being a very sensitive spectrometer SAFARI reveals the physical phenomena occurring within an obscured environment.  SAFARI opens the window to the very distant Universe as well as nearby star and planet formation.

Calibration of SAFARI is challenging.  SAFARI is exceptionally sensitive, but planets and stellar calibrators within the detection range of earlier far-infrared space observatories will easily saturate the very sensitive detectors.  At 8K, the cold primary will have emission only at the longest wavelengths.  A deep integration on a distant galaxy will detect the local foreground emission of the zodiacal dust cloud and the Galactic emission which must be removed.  Observing faint line emission on bright continua from nearby proto-planetary disks will likely show non-linearity effects.

We outline the key components necessary to calibrate the SAFARI instrument.  This article is organized as follows:  section \ref{sec:description} describes the SPICA mission and SAFARI instrument.  In section \ref{sec:obsmodes} we present a conceptual operation of SAFARI.  Section \ref{sec:requirements} describes the calibration requirements and section \ref{sec:strategy} outlines how those requirements could be met. A summary of the main points is presented in section \ref{sec:summary}. 

\section{SAFARI DESCRIPTION}
\label{sec:description}
%Reminder to not forget this!
%This section will describe SAFARI 3.0.  It should indicate the layout of the spaxels and where the calibration source and beam steering mirror are in the optical chain. 
%SAFARI 3.0  4 bands from 34 to 230 microns.  Each band 147 spectral pixels at 5 spatial positions.   Total number of TES detectors 4*5*147, 2940.  Show an image of the focal plane layout of one grating module.

% start of Edgar Section

\begin{figure} [ht]
   \begin{center}
   \begin{tabular}{c} %% tabular useful for creating an array of images 
   \includegraphics[width=12cm]{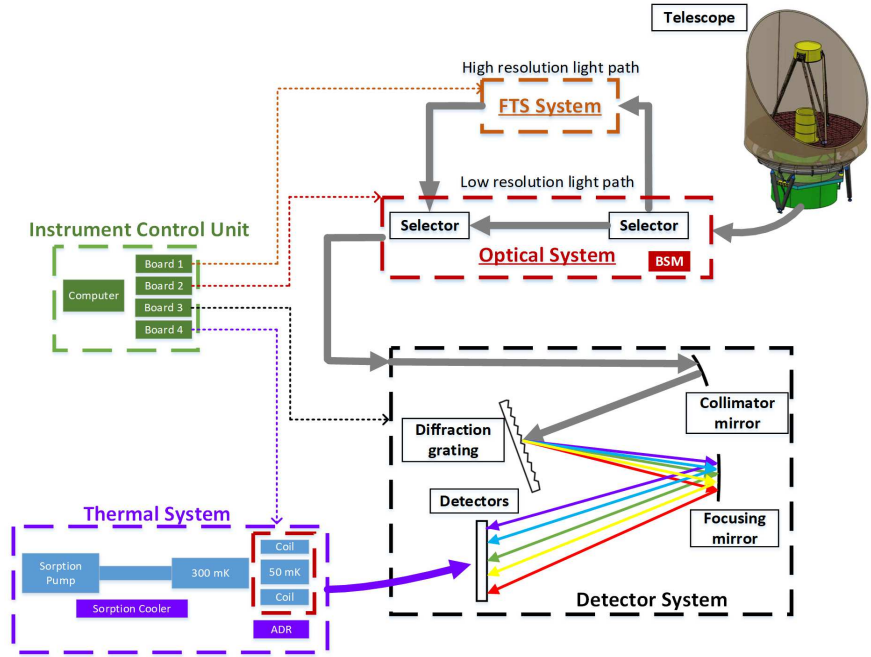}
   \end{tabular}
   \end{center}
   \caption[design] 
%>>>> use \label inside caption to get Fig. number with \ref{}
   { \label{fig:design} 
Conceptual design of SAFARI. A selectable low/high resolution spectrograph/spectrometer is proposed to comply with the two different resolutions required by key SAFARI science questions. The beam steering mirror (BSM) together with the mode selector and the calibration source (not shown) make up the optical system.}
   \end{figure}

The SPICA Far Infrared Spectrometer (SAFARI)\cite{2020..SAFARI.R} is a dual resolution grating spectrometer, combined with a high resolution post dispersed Fourier Transform Spectrometer. A simplified version of the conceptual design can be seen in Figure \ref{fig:design}.  The design is simple: an optical system selects if the light beam from the telescope is directly coupled to a detector system (low resolution light path, LR), or the beam is first directed to a Fourier Transform Spectrometer (high resolution light path, HR).  The mode selector is the mechanisms which controls which path is used.

The LR sends the light to a spectrometer with diffraction gratings as dispersing elements. Diffraction gratings were selected to minimize the loss of light in the materials due absorption and dispersion. The gratings are designed to provide an optical resolution in the spectral axis of 250. In the HR, the light passes first through a Fourier Transform Spectrometer, then through the grating spectrograph. This post dispersion operation increases the spectral resolution up to the range between 1500 to 12000, depending on the wavelength range (34 $\mu$m to 230 $\mu$m) and also reduces the background photon noise by dividing the incoming light into all detectors.

An important component of the optical system is the beam steering mirror (BSM). It has two main functions, one is to direct the incoming radiation from the telescope to a specific point in the focal plane, the other is to direct the radiation from an internal calibration source to the focal plane. The focal plane has a $2 \times 2 $ square arcminutes field of view (FOV). 

Because of the high bandwidth required, along with absolute accuracy and uniformity along the whole band in both intensity and spectral measurements, it was decided to divide the spectral range in 4 sections, one for each of the 4 grating modules. In this way, four different optimizations are possible in optical performance, mass, volume and accommodation in the available space on SPICA.

%\textcolor{red}{Describe a grating module to link the next section...}

% end of Edgar Section

\begin{figure} [ht]
   \begin{center}
   \begin{tabular}{c} %% tabular useful for creating an array of images 
   BSM position 1 \\
   \includegraphics[width=9cm]{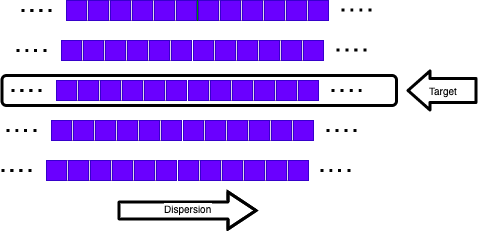} \\
   BSM position 2 \\
    \includegraphics[width=9cm]{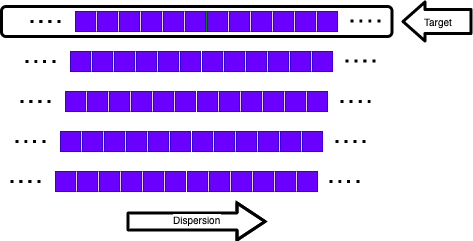}
   \end{tabular}
   \end{center}
   \caption[positions] 
%>>>> use \label inside caption to get Fig. number with \ref{}
   { \label{fig:layout} 
Layout of detector arrays in one grating module.  Each detector is shown as a blue rectangle.  A row of pixels make up a spaxel.  Detector rows probe a spectral segment for 5 positions offset in the sky.  The detector row (spaxel) centered on the target in the sky is highlighted with a box.

Note, all spaxels are receiving sky background radiation, only the central spaxel is receiving radiation from the target.}
   \end{figure}

\begin{figure} [ht]
   \begin{center}
   \begin{tabular}{c} %% tabular useful for creating an array of images 
   
   \includegraphics[width=9cm]{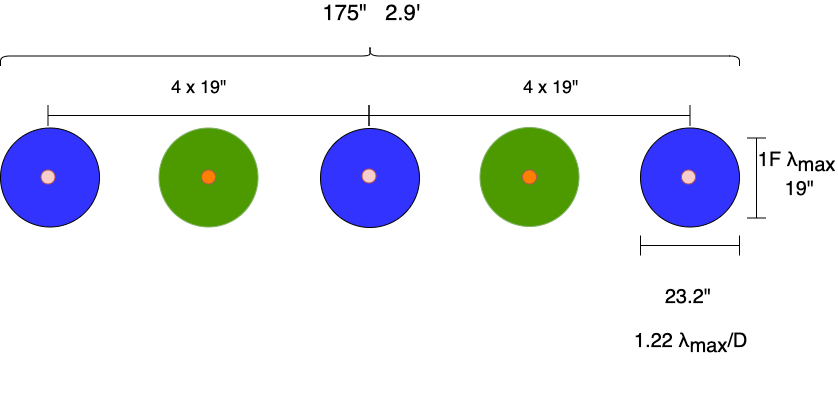} 
   \end{tabular}
   \end{center}
   \caption[positions] 
%>>>> use \label inside caption to get Fig. number with \ref{}
   { \label{fig:fivespaxel} 
Focal plane footprint on sky. This figure shows the full width at half maximum of the shortest (center of each circle) and longest wavelength (large circles) pixels of the 5 spaxels assuming Gaussian beams at each wavelength.
}
   \end{figure}

In each of the 4 grating modules the dispersed optical signal is recorded by 5 arrays of TES bolometers with a sensitivity close to background-limited \cite{SAFARI_det}.  The 5 arrays record spatially different signals from the sky but are co-aligned across the grating modules.   Each array of 147 individual bolometers constitutes a spaxel, each spectral element (bolometer) in the spaxel is a spectral pixel.  Figure \ref{fig:layout} demonstrates the layout and the steps of an observation.  A single spaxel produces a spectrum sampled once per resolution element.  To achieve two samples per spectral resolution element (Nyquist sampling), the arrays are offset by a fraction of a spectral resolution element.  For example spaxels 1, 3 and 5 would be offset by 1/2 a resolution element.  A beam steering mirror (BSM) is used to direct the input signal from a given sky direction onto a desired spaxel.  Modulating the signal with the BSM between spaxel 3 and 1 (or 3 and 5), would allow for a fully sampled spectrum to be recorded.   In Figure \ref{fig:layout}, the BSM steers the beam from the target onto spaxel 3 and then spaxel 1 and back.  Thus modulating the signal.  The other spaxels receive signal near but off the target which is taken to be background.  As seen in Figure \ref{fig:fivespaxel}, the background is measured at the same time as the target at positions between about 40 to 100 arcseconds off target.

Grating modules are co-aligned on the sky, spaxel 1 of grating module 1 sees the same sky position as spaxel 1 of grating modules 2,3 and 4.  The final product of SAFARI are fully sampled spectra from 34 to 230 microns.  Figure \ref{fig:fivespaxel} shows the footprint of SAFARI on the sky.  The circles represent the SAFARI beams (in full width at half maximum, FWHM) from 34 $\mu m$ to 230 $\mu m$.  This figure shows the spatial offset between spaxels.  Each spaxel has $\approx$ 40 arcsec separation from the next.

%%\subsubsection{TES bolometers}
%SAFARI makes use of transition edge superconducting (TES) bolometers .  These devices have a linear response over a high dynamic range\cite{Lee:98}\cite{2012RScI...83g3113D}, furthermore.  

\section{OBSERVING MODES}
\label{sec:obsmodes}
SAFARI measures spectra with either low resolution gratings (LR mode) or high resolution (HR mode) by inserting an FTS in front of the gratings (see Figure \ref{fig:design}).

\subsection{Low resolution single spectra}
Consider an emission source with spectral flux density $S_\nu$, an instrument response function $F_\nu$ and a spaxel dependent aperture efficiency $\eta_j$.  

Assuming a linear detector's response, the measured flux density of a source in LR mode over the pixel $i$ is:
\begin{equation}
    S_{ij} = \frac{{\int_{\Delta \nu_{ij}} S_\nu F_\nu \eta_j d\nu}}{{\int_{\Delta \nu_{ij}} F_\nu \eta_j d\nu}} 
\end{equation}

where the integral is over the bandwidth $\Delta \nu$ of the $i^{th}$ pixel of spaxel {j}.  Each pixels sees a bandwidth $\Delta \nu$ in either the LR mode or the HR mode.  

LR spectra of a single objects are a major scientific mode for SAFARI.  The purpose of this mode is to measure fully sampled low resolution spectrum (near ${\lambda} / {\Delta \lambda} = 250$).  Begin  background limited, One of SAFARI's main science goals is to measure the interstellar medium (ISM) in distant galaxies.  

SAFARI is designed to always integrate on source.  The BSM places the target on one spaxel,  while the other 4 spaxels see nearby (40-100 arcseconds) off background.  The BSM then switches the target to a different spaxel.  Thus an integration is a modulation between target and nearby background. In Figure \ref{fig:layout} when the BSM is in position one, the target emission falls on the middle spaxel, when the BSM shifts to position two the middle spaxel receives radiation from the background.  

The spaxels will be offset from each other also in the dispersion direction by a fraction of a spectral resolution element. With spectrally offset spaxels (1/2 a spectral resolution element), the final spectrum is created by interlacing the integrations from two different spaxels. 
Due to construction of a linear TES array within a grating module there will be gaps in the spectral coverage between TES bolometers.  The interlacing also fills these spectral gaps.

Each pixel is modulated by chopping between an ON target position and an OFF target position.  The power measured within a pixel-bandwidth, $\Delta \nu_i$, while ON target, $P_{i,ON}$, is a combination of power from the target in $\Delta \nu_i$, the power from the foreground emission in $\Delta \nu_i$  and the "power" registered with no loading due to the bias voltage of the pixel, $P_{i,0}$:
\begin{equation}
  P_{ij,ON} = P_{ij,target} +P_{ij,bkg^a} + P_{ij,0}  \label{equ:on}
\end{equation}
\begin{equation}
  P_{ij,OFF} = P_{ij,bkg^b} + P_{ij,0}  \label{equ:off}
\end{equation}
The subscript $ij$ refers to the $i^{th}$ pixel of spaxel $j$. 
Both $P_{ij,bkg^a}$ and $P_{ij,bkg^b}$ are due to foreground emission. 
As can be seen from Figure \ref{fig:fivespaxel}, the background positions are separated by about 80 arcseconds on the sky.  For most diffuse background sources (Zodiacal emission or Galactic dust emission) there will not be any significant spatial variation over within a few arcminutes, hence $P_{ij,bkg^a} \approx P_{ij,bkg^b}$.  
In case the background emission is highly variable,  the chopping can be performed more symmetrically by making use of more spaxels (chopping from spaxel 3 to 1 then 3 to 5) and interpolating to determine the background emission $P_{bkg^a}$.  Furthermore, the background emission is broadband continuum emission and nearby pixels within a spaxel at different wavelengths can also be used to determine the background power. The power from the target is:
\begin{equation}
  \Delta P_{ij} = P_{ij,ON}-P_{ij,OFF} = P_{ij,target} + P_{ij,bkg^a} - P_{ij,bkg^b} \approx P_{ij,target} \label{equ:diff}
\end{equation}
However, as stated earlier, the spectrum of the target should be sampled twice. The modulation steps in represented in Equation \ref{equ:diff} observe the source only once within the bandwidth of the pixel.  We therefore make use of the spaxel which is chopped onto the the position of the target $j + 2$ or $j-2$.
\begin{equation}
  P_{ij \pm 2,ON} = P_{ij \pm 2,target} +P_{ij \pm 2,bkg^a} + P_{ij \pm 2,0}  \label{equ:2ndspaxelon}
\end{equation}
Here again $i$ refers to the $i^th$ pixel of spaxel $j+2$ or $j-2$.
\begin{equation}
  \Delta P_{ij\pm 2} = P_{ij\pm 2,target} \label{equ:diff2ndspaxel}
\end{equation}
Interleave the signals of pixel $i$ from the different spaxels gives the final spectrum:
\begin{equation}
  S_\lambda =I(\lambda)[P_{ij,target},P_{ij\pm 2,target}] \label{equ:LRspectrum}
\end{equation}
Where $I(\lambda)$ reconstructs (interpolates) the interleaved target signal.

\subsubsection{Low resolution mapping}
%\textcolor{red}{Although mapping for the Yellow book was always identified as raster mapping.  I think it would be good here to explore OTF mapping as well.  Specifically the magic angles to provide full spatial Nyquist sampling.}
Some of SAFARI's key science questions also call for a spectral mapping mode.  Small raster maps could be made using the pointed mode described above but at different positions on the sky.  The BSM can cover a 2'x2' field of view to produce a small map.  Larger maps could be made by stitching smaller maps together.  A more efficient On-The-Fly mapping  scheme is also possible.  In this mode, SAFARI is integrating while the telescope slews.

\subsection{High resolution single spectra}
As described in section \ref{sec:description}, the mode selector is used to redirect the telescope beam through the FTS in high resolution mode (HR).
In the HR mode, each detector produces an interferogram $S_{ij}(t)$  during the scanning of the FTS mechanism over time $t$, such that the Fourier Transform produces the spectrum of the source :
\begin{equation}
    S_{ij}(\nu) = FT(S_{ij}(t))  \label{equ:HR}
\end{equation}

The FTS mechanism is then scanned its length and back.  
%As the FTS mirror scans, an interferogram is then passed through the grating spectrometer onto each pixel in the bandwidth set by the grating. Bandwidth of each pixel is near ${\lambda} / {\Delta \lambda} = 250$.  
Given 4 grating modules each with 5 spaxels of roughly 150 pixel, a single FTS scan will produce roughly 3000 interferograms.

If more than one scan is required for the proper integration, either the BSM or a re-pointing of the satellite can be used to place the target on a different spaxel to address the small band gaps between pixels.

\section{CALIBRATION REQUIREMENTS}
\label{sec:requirements}
When linear, the voltage modulation of pixel $i$ in spaxel $j$ is directly related to the absorbed power:
\begin{equation}
\label{equ:maincal}
P_{ij} = K_{ij} \Delta V_{ij}
\end{equation}
Where $K_{ij}$  depends pixel, spaxel, BSM position and potentially if the source is extended or point-like.

\subsection{Calibration items}
Table \ref{tab:calitems} lists the calibration items that are needed to address the requirements.  The table identifies the dependencies of the calibration item on wavelength ($\lambda$),  pixel ($i$), spaxel ($j$) or BSM position ($x$).   The BSM allows for different optical paths through the SAFARI optics which could have a calibration impact. The BSM  also provides a position on the primary which may be at a slightly different temperature.

\begin{table}[ht]
\caption{Calibration items which follow from requirements and dependency for pixel $i$, spaxel $j$ } 
\label{tab:calitems}
\begin{center}       
\begin{tabular}{|l|l|l|l|} 
\hline
\hline
Item name & Item label & pixel, spaxel, BSM & Requirement \% \\
\hline
%\hline
Wavelength calibration & $\lambda$ & $i,j$  & 1.0 \\
\hline
Flat-field & FF & $i,j,x$ & 0.1   \\
\hline
Relative flux calibration & RSRF & $i,j,\lambda$ & 1.0   \\
\hline
Absolute flux Calibration & $F_{cal}$ & $\lambda$ & 10 \\
\hline
Cross talk & $X_{talk}$ & $i,j$ &  $<$ 0.1 \\
%\hline
%Overall efficiency & & & 10 \\
\hline

\end{tabular}
\end{center} 
\end{table}

\subsection{Wavelength/frequency calibration}
\label{sec:HRwavelenghtcal}
In LR mode, instrument level testing (ILT) provides most of the wavelength calibration.  Both the grating and the pixels are stationary.  This can be determined in laboratory testing and is not expected to change.  However the calibration should be confirmed in flight.   

In HR mode determining the frequency calibration is more involved.  The high resolution is achieved with an FTS where the optical path difference (OPD) creates an iterferogram. The Fourier Transform of that interferogram is the source signal.  The OPD must be accurately determined at all time steps during the scan of the FTS mechanism.  One key component is the zero path difference (ZPD) when there is no difference in the path lengths between the two arms.  There will be almost 3000 pixels which can be used to help determine ZPD.

\subsection{Spectral line-shape}
While not a specific calibration requirement, spectral line-shapes should be determined across the frequency range of SAFARI.  In the LR mode, a spectral line is observed by two spaxels and spread over multiple pixels.  The observation over different spaxels requires interlacing of the final spectrum and proper interpolation onto a finer wavelength grid.  The line-shape will be measured towards external sources with unresolved lines and compared with the instrument simulator \cite{2020.Simulator}.  

The spectral lines are observed and constructed via signals from slightly paths through the optics:  the BSM is used to redirect the source signal to a different spaxel (see Figure \ref{fig:layout}).  Any BSM dependence on the response will have to be measured/modelled. 

In the HR mode, the spectral line shape is determined by the spectral response of the pixel itself.  Again, the comparison with the instrument model should be made to raise confidence that the instrument and response are well understood.  

SAFARI will observe brighter nearby proto-planetary disks. For these sources, the line shape will be dependent on correcting any non-linear response of the detectors while making the interferogram.  A proper measurement and monitoring of the non-linear response of the detectors will be needed.  

\subsection{Relative flux calibration}
The relative flux calibration refers to the flux calibration of each pixel relative to the other pixels.  For pixels close in wavelength %(within \textcolor{red}{TBD $\mu m$})
, the requirement is 0.1\% in LR mode.  IN HR the requirement is 1\%.

The relative response functions of the detectors will be mapped during ILT.  The ILT measurements are required throughout the mission for proper calibration. ILT measurements provide the spectral/spatial of each bolometer.  

The internal calibration source provides a stable reference for extended emission.
It will not provide information regarding point source coupling.  But it will aid in monitoring SAFARI as a whole.

In flight, the RSRF will be verified using calibration standards,  both stellar and broadband continuum sources.

\subsubsection{Low resolution RSRF}
The main calibration parameter for the relative flux calibration is to determine the response relative to adjacent pixels.  In this sense the low resolution RSRF is more akin to a flat-field.   An internal calibration source that provides a stable featureless extended signal will be key for determining and monitoring the flat-field (see Section \ref{sec:CSA}).  Broadband continuum point sources will be needed to determine a point source correction to the flat-field.  Detailed modelling of SAFARI is necessary to determine if there will be a position dependence and how sensitive to pointing errors the flat-field/RSRF will be.   \cite{2020.Simulator}.

\subsubsection{High resolution RSRF}
In HR mode, a major calibration step will be the creation of a "clean" interferogram before Fourier transforming\cite{2016MNRAS.458.1977F} \cite{2001ftsp.book.....D}.  Each pixel creates an interferogram.  This provides redundancy and an opportunity to examine the interferograms for instrumental correlations which otherwise would impact the quality of the data. 

Within the bandwidth of a pixel, a properly corrected interferogram has the most impact on the relative response.  Broad bandwidth FTS spectrometers like SPIRE required special processing steps to deal with the changing beam size \cite{2013A&A...556A.116W}, the bandwidth over a grating pixel is small enough not to show beam size issues between pixels.  Observations of broadband sources like asteroids should provide both RSRF and absolute calibrations.   

Monitoring of the RSRF in high resolution will be performed with observations of the internal calibration source (see Section \ref{sec:CSA}). 

The RSRF in high resolution mode is a measure of the response at frequencies within the bandpass of each pixel. 
The determining of the "flat-field" as described above will remain import for proper spectral stitching across the entire SAFARI wavelength range.

\subsection{Absolute flux calibration}
Absolute flux calibration will rely on observing well modelled celestial sources.  Traditionally, these sources have been a combination of stars, planets and minor planets.  Unfortunately, Uranus and Neptune have flux densities greater than a few hundred Jy in SAFARI wavelength range and are all too bright (Figure \ref{fig:saturation}). 

Herschel observed stars, planets and minor planets as flux calibrators \cite{2014ExA....37..253M}.  Sources such as $\gamma Dra$ are observable with SAFARI as are asteroids and minor planets. 

Figure \ref{fig:saturation}, shows a top level overview of possible calibrators.  The figure identifies two limits.  The top hatched area identifies at which flux density level SAFARI will likely saturate.  High resolution observations saturate at roughly twice the limit.  The lower solid line indicates the flux level that can be achieved in a 10 minute observation ($5\sigma$).  To be a viable calibration source, the object should observable in a relatively short amount of time.  The dimmest star in the diagram, $\gamma$ Dra, makes it clear that no single star is usable across the entire SAFARI range, but some asteroids would be.

\section{CALIBRATION STRATEGY}
\label{sec:strategy}
Calibration activities take place on different time scales.  The observing mode must first and foremost guarantee that integrating on a source will reduce the observed noise.  This is achieved by periodically monitoring the system gain by flashes of the calibration source. Data from these flashes will be used to bring every observation to a standard level. This is not an absolute flux calibration, but a link between normal operations and the absolute flux calibrators described below.

Another requirement from the system is to produce a properly sampled spectrum.  Properly sampled fulfills the Nyquist sampling requirement of at least two samples per resolution bandwidth. In the low resolution mode of grating data, this is accomplished by the spaxel arrays being offset from each other by a fraction of a spectral resolution element.  This is shown in Figure \ref{fig:layout}.  In high resolution mode the FTS fulfills the requirement by design.

\subsection{Pre-flight characterization}
\label{sec:ILT}
During instrument level testing (ILT), many calibration parameters will be determined.  Among these will be the actual layout of the focal plane as-built.  In combination with the grating, this layout determines the wavelength calibration.  

ILT will also map the focal plane in the spatial and spectral dimensions.  This map results in the relative spectral response (RSRF) for each detector in every spaxel of all four grating modules.  When combined with a telescope model, the RSRF is the instrument response to incoming radiation. This will result in roughly 3000 response functions.  
Mapping of the focal plane should also include checks of how impact of the BSM on the RSRF and whether any corrections are needed when combining (interlacing) signals from two different BSM positions.  

Another outcome of the RSRF mapping above will be a cross talk matrix.  Likely, there will be cross talk given the frequency domain multiplexing.  Depending on the layout of the multiplexing (all within a spaxel or across spaxels), identifying cross-talk in-flight could be very difficult.  A laboratory measurement will be necessary.   
Specifically important for bright sources will be determining the non-linear response.  Over a wide dynamic range, the detector response is linear, but timing effects appear once the source brightness approaches saturation.  Being intrinsic to the detector/readout system, the non-linearity is not expected to change once measured.  Although that should be verified in-flight.

One final ILT activity will be to determine the stability of various components of the SAFARI instrument.  Higher stability elements require less frequent calibration activities and increase the overall instrument efficiency.  The stability analysis will specify the cadence of calibrations.

\subsection{Internal calibration source}
\label{sec:CSA}
The internal calibration source is present to provide a stable consistent signal across all SAFARI wavelengths.  It can be used to monitor the overall response of the system as well as monitor the response gain by providing signals which modulate between two power levels.  Such a "flash" signal was key to the SPIRE response monitoring and calibration scheme for both their photometers and FTS spectrometer \cite{2013MNRAS.433.3062B}, \cite{2014MNRAS.440.3658S}.  The current design of the calibration source makes use of two micro-lamps in an integration sphere.  These lamps produce black-body radiation at two different temperatures.  The signal produced from the two lamps is a broad, featureless spectrum covering the entire range of SAFARI and illuminating all 5 spaxels.  

The useful time scale of the calibration source is per observation to daily.  As a monitoring component, the internal source will be used to bring the entire SAFARI response to a standard time in order to bootstrap the absolution calibration to the entire mission.

The SAFARI internal calibration source will need to have the following capabilities \cite{2020..CSA.}.

\begin{itemize}
    \item Standard spectroscopic source signal.
    This should be a flat spectral signal that illuminates the entire wavelength range of SAFARI as well as the 5 spaxels.  This signal can be used to monitor the relative response of SAFARI in HR mode.
    \item Small modulated signal.  This is also a flat signal covering the entire SAFARI range, but can be modulated between two levels.  This signal will be used to monitor the gain of the detector system in both LR and HR modes.  In HR mode, the FTS mechanism will have to be set at ZPD and then modulate the calibration source (see Section \ref{sec:HRwavelenghtcal}).
    \item Dark signal.  When off, the calibration source will be at the 4K and provide a dark reference signal.
    \item Comb spectrum.  In a separate mode, the calibration source can provide a comb signal for frequency calibration of the FTS.  This will not only be useful for frequency calibration of the FTS, but will also contribute to the wavelength calibration in low resolution  mode since the FTS requires the grating to disperse the signal into finer bandpasses. 
    \item Saturating source possibility.  For monitoring purposes, one potential capability for the calibration source would be to produce a signal which partially saturates the instrument.  This provides a check over time of any fundamental changes in the instrument.
\end{itemize}

\subsection{External calibrators}
External calibrators are needed for wavelength calibration, line shape, relative and absolute flux calibration.

\begin{figure} [ht]
   \begin{center}
   \begin{tabular}{c} %% tabular useful for creating an array of images 
   \includegraphics[width=14cm]{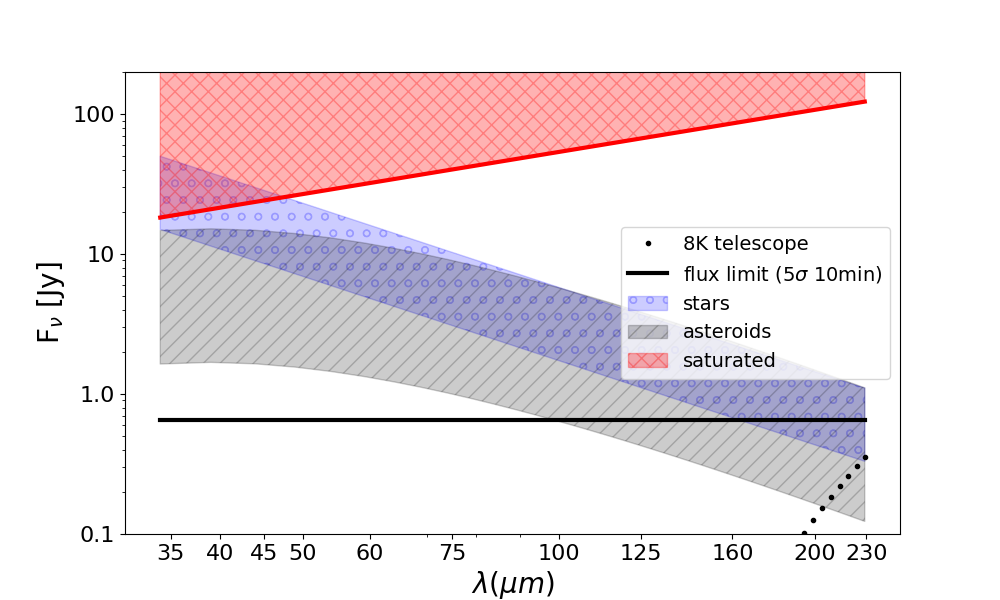}
   \end{tabular}
   \end{center}
   \caption[calibrators] 
%>>>> use \label inside caption to get Fig. number with \ref{}
   { \label{fig:saturation} 
External calibrators. SAFARI can make use of some standard absolute flux calibrators.  It would be good to expand the potential list.  The saturation level is identified as the red square-hatch area.  Neptune and Uranus fall into this region as does Ceres.  The stellar calibration shows a range (in blue circles) of standard established calibrators from $\gamma$ Dra to $\alpha$ Boo.  The asteroid range shown in grey hatches is within a factor of 3 of the faintest asteroid calibrator considered for Herschel:  Lutetia \cite{2014ExA....37..253M}.  }
   \end{figure}

To verify the wavelength calibration, SAFARI will observe external narrow line sources such as planetary nebulae (PN).  The key here will be to find sufficiently weak PN that will not saturate SAFARI but bright enough to be observed in a reasonable time (see Figure \ref{fig:saturation}).  Being solely determined by the focal plane layout, the wavelength calibration is not expected to vary over the mission.  This still should be measured.  Observations of narrow line sources like PN will also determine, in-flight, the spectral line shape for both LR and HR modes.  

Figure \ref{fig:saturation} shows the flux ranges where external flux calibrators will be used.  The top boundary indicates where SAFARI would be saturated.  The black line indicates the $5\sigma$ flux limit in LR mode for a 10 minute integration. To keep calibration time to a minimum and maximize efficiency, integration times on calibration sources should be relatively short.  The lower bound for stars is based on the star $\gamma Dra$.  As can bee seen this will not saturate, but at long wavelengths, the integration time will be a concern.  In general stellar sources vary significantly across the SAFARI range making them less than ideal.  Asteroids are potentially a better option, since they can be proper point sources with flat spectra.  In the coming years effort should be invested in understanding these sources and improving the models.

\section{SUMMARY}
\label{sec:summary}
In this note the calibration requirements and strategy for the SAFARI instrument have been discussed. The main conclusions are as follows:
\begin{itemize}
    \item SPICA/SAFARI will have roughly 3000  distinct spectral elements.  Each one requires spectral, spatial and non-linearity characterisation.  
    \item Many of the characterisation will be performed during ILT.  This will help confirm instrument simulations.
    \item SAFARI will make use of a very simple but highly repeatable internal calibration source, not for absolute calibration but rather to monitor the performance.
    \item The extraordinary sensitivity of the SAFARI's TES detectors make the instrument susceptible to saturation from standard established external calibrators such as planets and many stars.  
    \item The broad simultaneous wavelength range of SAFARI makes it challenging to find external calibrators which have sufficient flux at all wavelengths but do not saturate.
    \item In the coming years, the list of external calibrators in the far infrared needs to be expanded.  Likely objects will be asteroids, but also more distant planetary nebula and AGB stars. The stellar flux calibrator network for JWST - extending mid-infrared calibration networks to fainter targets - is a promising starting point. 
\end{itemize}

The next generation of far infrared instruments will make use of the technological advances in detector technology of the past decades. The advances have been significant and have substantially increased sensitivity, linearity and dynamic range.  The SPICA/SAFARI instrument would have been one of these major advances by taking advantage of a cold large telescope and the most sensitive detectors.  SAFARI does not image the far infrared sky but rather measures in exquisite detail the rich far infrared spectrum.

\acknowledgments % equivalent to \section*{ACKNOWLEDGMENTS}       
 
R.S. thanks David Naylor for many useful discussions and his insight to the needs of an FTS.  
A.L. acknowledges the support from Comunidad de Madrid through the Atracción de Talento grant 2017-T1/TIC-5213. This research has been partially funded by the Spanish State Research Agency (AEI) Project MDM-2017-0737 Unidad de Excelencia “María de Maeztu”- Centro de Astrobiología (INTA-CSIC).

% References
\bibliography{safarical} % bibliography data in report.bib

\begin{thebibliography}{10}

\bibitem{2020..SPICA.R}
Roelfsema, P.~R., Shibai, H., Kaneda, H., Sauvage, M., Najarro, F., Bradford,
  C.~M., Yamada, T., Tauber, J., Audard, M., Doi, P. D.~Y., Elbaz, D., Griffin,
  M., Helmich, F., Heske, A., Honda, M., Jellema, W., Juvela, M., Kamp, I.,
  Kerschbaum, F., Kiss, C., Kohno, K., Krause, O., de~Lange, G., Larsson, B.,
  Martín-Pintado, J., Hideo~Matsuhara, T.~N., Nagao, T., Naylor, D., Nomura,
  H., Ogawa, H., Onaka, T., Rouquet, J., Spinoglio, L., Szczerba, R., van~der
  Tak, F., Vandenbussche, B., and Wang, S.-Y., ``The joint infrared space
  observatory spica: Unveiling the obscured universe,'' in [{\em Space
  Telescopes and Instrumentation 2020: Optical, Infrared, and Millimeter Wave
  (Conference 11443)}{\nolinebreak\hspace{0.1em}]},  {\em Proc. SPIE} {\bf
  11443}(11443-205) (2020).

\bibitem{SAFARI_det}
Audley, M.~D., de~Lange, G., Gao, J.-R., Jackson, B.~D., Hijmering, R.~A.,
  Ridder, M.~L., Bruijn, M.~P., Roelfsema, P.~R., Ade, P. A.~R., Withington,
  S., Bradford, C.~M., and Trappe, N.~A., ``The safari detector system,'' in
  [{\em Milimeter, Submillimeter and Far-Infrared Detectors and Instrumentation
  for Astronomy IX)}{\nolinebreak\hspace{0.1em}]},  {\em Proc. SPIE} {\bf
  107080}(107080K) (2018).

\bibitem{2020..SAFARI.R}
Roelfsema, P.~R., Dieleman, P., Jellema, W., de~Lange, G., Evers, J., Giard,
  M., Najarro, F., Bradford, C.~M., Audard, M., Doi, Y., Griffin, M., Helmich,
  F., Juvela, M., Kerschbaum, F., Kiss, C., Krause, O., Larsson, B., Naylor,
  D., Spinoglio, L., Szczerba, R., van~der Tak, F., Vandenbussche, B., and
  Wang, S.-Y., ``The safari grating spectrometer for spica: Extreme
  spectroscopic sensitivity in the far-ir,'' in [{\em Space Telescopes and
  Instrumentation 2020: Optical, Infrared, and Millimeter Wave (Conference
  11443)}{\nolinebreak\hspace{0.1em}]},  {\em Proc. SPIE} {\bf
  11443}(11443-231) (2020).

\bibitem{2020.Simulator}
Jellema, W., Spinoglio, L., Pascale, E., Savini, G., Fernandez-Ontiveros, J.,
  Causi, G.~L., Shipman, R., Griffin, M., Naylor, D., and van Loon, D., ``The
  spica-safari instrument and scientific performance simulator,'' in [{\em
  Space Telescopes and Instrumentation 2020: Optical, Infrared, and Millimeter
  Wave (Conference 11443)}{\nolinebreak\hspace{0.1em}]},  {\em Proc. SPIE} {\bf
  11443}(11443-225) (2020).

\bibitem{2016MNRAS.458.1977F}
{Fulton}, T., {Naylor}, D.~A., {Polehampton}, E.~T., {Valtchanov}, I.,
  {Hopwood}, R., {Lu}, N., {Baluteau}, J.~P., {Mainetti}, G., {Pearson}, C.,
  {Papageorgiou}, A., {Guest}, S., {Zhang}, L., {Imhof}, P., {Swinyard}, B.~M.,
  {Griffin}, M.~J., and {Lim}, T.~L., ``{The data processing pipeline for the
  Herschel SPIRE Fourier Transform Spectrometer},'' {\em Monthly Notices Royal
  Astronomical Society}~{\bf 458},  1977--1989 (May 2016).

\bibitem{2001ftsp.book.....D}
{Davis}, S.~P., {Abrams}, M.~C., and {Brault}, J.~W.,  [{\em {Fourier transform
  spectrometry}}{\nolinebreak\hspace{0.1em}]}, Academic Press (2001).

\bibitem{2013A&A...556A.116W}
{Wu}, R., {Polehampton}, E.~T., {Etxaluze}, M., {Makiwa}, G., {Naylor}, D.~A.,
  {Salji}, C., {Swinyard}, B.~M., {Ferlet}, M., {van der Wiel}, M.~H.~D.,
  {Smith}, A.~J., {Fulton}, T., {Griffin}, M.~J., {Baluteau}, J.~P.,
  {Benielli}, D., {Glenn}, J., {Hopwood}, R., {Imhof}, P., {Lim}, T., {Lu}, N.,
  {Panuzzo}, P., {Pearson}, C., {Sidher}, S., and {Valtchanov}, I.,
  ``{Observing extended sources with the Herschel SPIRE Fourier Transform
  Spectrometer},'' {\em Astronomy and Astrophysics}~{\bf 556},  A116 (Aug.
  2013).

\bibitem{2014ExA....37..253M}
{M{\"u}ller}, T., {Balog}, Z., {Nielbock}, M., {Lim}, T., {Teyssier}, D.,
  {Olberg}, M., {Klaas}, U., {Linz}, H., {Altieri}, B., {Pearson}, C., {Bendo},
  G., and {Vilenius}, E., ``{Herschel celestial calibration sources. Four large
  main-belt asteroids as prime flux calibrators for the far-IR/sub-mm range},''
  {\em Experimental Astronomy}~{\bf 37},  253--330 (July 2014).

\bibitem{2013MNRAS.433.3062B}
{Bendo}, G.~J., {Griffin}, M.~J., {Bock}, J.~J., {Conversi}, L., {Dowell},
  C.~D., {Lim}, T., {Lu}, N., {North}, C.~E., {Papageorgiou}, A., {Pearson},
  C.~P., {Pohlen}, M., {Polehampton}, E.~T., {Schulz}, B., {Shupe}, D.~L.,
  {Sibthorpe}, B., {Spencer}, L.~D., {Swinyard}, B.~M., {Valtchanov}, I., and
  {Xu}, C.~K., ``{Flux calibration of the Herschel SPIRE photometer},'' {\em
  Monthly Notices Royal Astronomical Society}~{\bf 433},  3062--3078 (Aug.
  2013).

\bibitem{2014MNRAS.440.3658S}
{Swinyard}, B.~M., {Polehampton}, E.~T., {Hopwood}, R., {Valtchanov}, I., {Lu},
  N., {Fulton}, T., {Benielli}, D., {Imhof}, P., {Marchili}, N., {Baluteau},
  J.~P., {Bendo}, G.~J., {Ferlet}, M., {Griffin}, M.~J., {Lim}, T.~L.,
  {Makiwa}, G., {Naylor}, D.~A., {Orton}, G.~S., {Papageorgiou}, A., {Pearson},
  C.~P., {Schulz}, B., {Sidher}, S.~D., {Spencer}, L.~D., {van der Wiel},
  M.~H.~D., and {Wu}, R., ``{Calibration of the Herschel SPIRE Fourier
  Transform Spectrometer},'' {\em Monthly Notices Royal Astronomical
  Society}~{\bf 440},  3658--3674 (June 2014).

\bibitem{2020..CSA.}
Wu, E. S.-Y.~W., Wang, M.-J., Chen, T.-J., Chang, Y.-P., Huang, Y.-R., Wang,
  C.-L., Chiu, C.-P., Pei, T.-H., Evers, J., Arrazola, D., Eggens, M., Martin,
  S., and Jellema, W., ``The calibration source assembly for spica/safari
  instrument,'' in [{\em Space Telescopes and Instrumentation 2020: Optical,
  Infrared, and Millimeter Wave (Conference
  11443)}{\nolinebreak\hspace{0.1em}]},  {\em Proc. SPIE} {\bf
  11443}(11443-216) (2020).

\end{thebibliography}
\bibliographystyle{spiebib} % makes bibtex use spiebib.bst

\end{document}